\documentclass[pra, 12pt, showpacs, showkeys, eqsecnum, nofootinbib]{revtex4}

\begin{document}
\title{Symmetry examples in open quantum dynamics}
\author{Thomas F. Jordan}
\email[email: ]{tjordan@d.umn.edu}
\affiliation{Physics Department, University of Minnesota, Duluth, Minnesota 55812}
\author{San Ha Seo}
\email[email: ]{seoxx124@d.umn.edu}
\affiliation{Physics Department, University of Minnesota, Duluth, Minnesota 55812} 

\begin{abstract}

Dependent symmetries, symmetries that depend on the situation of the subsystem in a larger closed system, are explored by looking at simple examples. This is a new kind of symmetry in the open quantum dynamics of a subsystem

Each symmetry implies a particular form for the results of the open dynamics. The forms exhibit the symmetries very simply. It is shown directly, without assuming anything about the symmetry, that the dynamics produces the form, but knowing the symmetry and the form it implies can reduce what needs to be done to work out the dynamics; pieces can be deduced from the symmetry rather that calculated from the dynamics.

Symmetries can be related to constants of the motion in new ways. A quantity might be a \textit{dependent constant of the motion}, constant only for particular situations of the subsystem in the larger system. In particular, a generator of dependent symmetries could represent a quantity that is a dependent constant of the motion for the same situations as for the symmetries. The examples present a variety of possibilities. Sometimes a generator of dependent symmetries does represent a dependent constant of the motion. Sometimes it does not. Sometimes no quantity is a dependent constant of the motion. Sometimes every quantity is.

\end{abstract}


\maketitle

\newpage

\section{Introduction}\label{one}

A broad new definition of symmetries in open quantum dynamics was introduced recently and explored by looking at examples.\cite{Jordan} It uses equations for physically meaningful numbers to reformulate and extend to open quantum dynamics the rule, familiar from the dynamics of an entire closed system, that a unitary symmetry operator commutes with the Hamiltonian.\cite{note} Many symmetries of open quantum dynamics, defined by this reformulated rule, are described by unitary symmetry operators that do not commute with the Hamiltonian for the dynamics of the entire closed system that produces the open dynamics of the subsystem. The examples worked out were mostly for independent symmetries, those that do not depend on correlations or absence of correlations between the subsystem and the rest of the entire system or on the state of the rest of the entire system.

Here we consider more examples of dependent symmetries, those that do depend on correlations or absence of correlations between the subsystem and the rest of the entire closed system or on the state of the rest of the entire system. The new definition is that a unitary operator $U$ describes a symmetry of the open dynamics of a subsystem $S$, for a Hamiltonian $H$ for the dynamics of the entire closed system, if \cite{Jordan}
\begin{equation}
\label{mvbasic}
\text{Tr} \left[We^{itH}U^{\dagger}QUe^{-itH}\right] = \text{Tr} \left[WU^{\dagger}e^{itH}Qe^{-itH}U\right]
\end{equation}
for any time $t$, for all the operators $Q$ for the physical quantities of $S$, and for the density matrices $W$ for the states of the entire system that describe particular correlations or absence of correlations between $S$ and the rest of the entire system and/or particular states of the rest of the entire system. If it is assumed that there are no correlations between $S$ and the rest of the entire system, this means that
\begin{equation}
\label{opbasicdep}
\text{Tr}_R \left[\rho_R e^{itH}U^{\dagger}QUe^{-itH}\right] = \text{Tr}_R \left[\rho_R U^{\dagger}e^{itH}Qe^{-itH}U\right]
\end{equation}
for any time $t$, for all the $Q$ for $S$, and for particular density matrices $\rho_R$ for the subsystem $R$ that is the rest of the entire system.

We work out examples of dependent symmetries in Section II. Each symmetry implies a particular form for the results of the open dynamics. For several examples, we lay out the form and can see the symmetry in it very simply. We do show directly, without assuming anything about the symmetry, that the dynamics produces the form. We also see that knowing the symmetry and the form it implies can reduce what needs to be done to work out the dynamics; pieces can be deduced from the symmetry rather that calculated from the dynamics. 

Oscillator examples described in Section II.C are translated into classical mechanics in Section II.C.4.

In Section III we ask whether the symmetries are related to constants of the motion. In keeping with the context of dependent symmetries, we consider that a quantity could be constant only for particular situations of $S$ in the larger system with $R$. If it is, we call it a \textit{dependent constant of the motion}. In particular, we can ask whether a generator of dependent symmetries represents a quantity that is a dependent constant of the motion for the same situations as for the symmetries. We find examples of a variety of possibilities. Sometimes a symmetry generator does represent a dependent constant of the motion. Sometimes it does not. Sometimes no quantity is a dependent constant of the motion. Sometimes every quantity is.

\section{Symmetry examples}\label{two}

For all these examples, we assume there are no correlations between the states of $S$ and $R$. The symmetries are just dependent on the state of $R$. The definition of symmetry is the statement that Eq.(\ref{opbasicdep}) holds for particular states of $R$.

\subsection{One qubit and one qubit} \label{qubit}

Let $S$ be a qubit described by Pauli matrices $\Sigma_1$, $\Sigma_2$, $\Sigma_3$ and $R$ a qubit described by Pauli matrices $\Xi_1$, $\Xi_2$, $\Xi_3$. 

\subsubsection{Hamiltonians $\gamma_1 \Sigma_1 \Xi_1 + \gamma_2 \Sigma_2 \Xi_2 + \gamma_3 \Sigma_3 \Xi_3 $} \label{Hfirst}

For our first examples, we consider Hamiltonians of the form
\begin{equation}
  \label{Hqd}
H = \frac{1}{2} \left[ \gamma_1 \Sigma_1 \Xi_1 + \gamma_2 \Sigma_2 \Xi_2 + \gamma_3 \Sigma_3 \Xi_3 \right]
\end{equation}
with real numbers $\gamma_1$, $\gamma_2$, $\gamma_3$. We use the previous calculation \cite{jordan05b,Jordan} that

\newpage

\begin{eqnarray}
  \label{USsu}
e^{itH}\Sigma_1 e^{-itH} & = & \Sigma_1 \cos \gamma_2 t\cos \gamma_3 t + \Xi_1 \sin \gamma_2 t\sin \gamma_3 t \nonumber \\
 && - \Sigma_2 \Xi_3 \cos \gamma_2 t\sin \gamma_3 t + \Sigma_3 \Xi_2 \sin \gamma_2 t\cos \gamma_3 t 
\end{eqnarray}
\begin{eqnarray}
 \label{US2}
e^{itH}\Sigma_2 e^{-itH} &=& \Sigma_2 \cos \gamma_3 t\cos \gamma_1 t + \Xi_2 \sin \gamma_3 t\sin \gamma_1 t \nonumber \\
 && - \Sigma_3 \Xi_1 \cos \gamma_3 t\sin \gamma_1 t + \Sigma_1 \Xi_3 \sin \gamma_3 t\cos \gamma_1 t, 
\end{eqnarray}
\begin{eqnarray}
  \label{US3}
e^{itH}\Sigma_3 e^{-itH}  &=& \Sigma_3 \cos \gamma_1 t\cos \gamma_2 t + \Xi_3 \sin \gamma_1 t\sin \gamma_2 t \nonumber \\
 && - \Sigma_1 \Xi_2 \cos \gamma_1 t\sin \gamma_2 t + \Sigma_2 \Xi_1 \sin \gamma_1 t\cos \gamma_2 t. 
\end{eqnarray}

The unitary operators
\begin{equation}
\label{rotR} 
U(u) = e^{-iu(1/2)\Sigma_3}
\end{equation}
change $\Sigma_1$ and $\Sigma_2$ to
\begin{eqnarray}
  \label{rot12}
U^{\dagger} \Sigma_1 U & = & \Sigma_1 \cos ut - \Sigma_2 \sin ut \nonumber \\
U^{\dagger} \Sigma_2 U & = & \Sigma_2 \cos ut + \Sigma_1 \sin ut 
\end{eqnarray}
and do not change $\Sigma_3$. They describe symmetries of the open dynamics of $S$ when
\begin{eqnarray}
  \label{USaa}
\text{Tr}_R\rho _R[e^{itH}\Sigma_1 e^{-itH}] & = & \Sigma_1 a(t) - \Sigma_2 b(t), \nonumber \\
\text{Tr}_R\rho _R[e^{itH}\Sigma_2 e^{-itH}] & = & \Sigma_2 a(t) + \Sigma_1 b(t), \nonumber \\
\text{Tr}_R\rho _R[e^{itH}\Sigma_3 e^{-itH}] & = & \Sigma_3 c(t) + d(t) 
\end{eqnarray}
with real functions $a(t)$, $b(t)$, $c(t)$, $d(t)$. We have already seen\cite{Jordan} and can easily check that this happens when $\gamma_1$ and $\gamma_2$ are equal and $\langle \Xi_1 \rangle $ and $\langle \Xi_2 \rangle $ are zero.

In particular cases, the form (\ref{USaa}) is simpler and accommodates more symmetries. If $\gamma_3$ is zero, then $b(t)$ is zero. If $\langle \Xi_3 \rangle $ is zero, so that $\langle \Xi_1 \rangle $, $\langle \Xi_2 \rangle $, $\langle \Xi_3 \rangle $ are all zero, then $b(t)$ and $d(t)$ are zero. In this latter case, there are additional symmetries described by all the unitary operators $U_R$ that are just for $R$ and do not depend on $S$. They imply\cite{Jordan} that the open dynamics of $S$ is the same for the state of $R$ represented by any $U_R\rho_R U_R^{\dagger}$ as it is for the state represented by $\rho_R $, so they imply that $b(t)$ and $d(t)$ are zero if it is assumed that just one of $\langle \Xi_1 \rangle $, $\langle \Xi_2 \rangle $, $\langle \Xi_3 \rangle $ is zero; the implications that can be taken from Eqs.(\ref{USsu})-(\ref{US3}) to make additional simplifications in the symmetric form (\ref{USaa}) when it is assumed that $\langle \Xi_1 \rangle $ is zero, for example, and the results of the dynamics are not changed by unitary transformations of the state of $R$ where $\langle \Xi_1 \rangle $ is zero to states where $\langle \Xi_2 \rangle $ or $\langle \Xi_3 \rangle $ is zero, are the same as the implications obtained when it is assumed that $\langle \Xi_1 \rangle $, $\langle \Xi_2 \rangle $, $\langle \Xi_3 \rangle $ are all zero.

Now, without assuming anything about $\gamma_1$, $\gamma_2$, $\gamma_3$ or $\langle \Xi_1 \rangle $, $\langle \Xi_2 \rangle $, $\langle \Xi_3 \rangle $, consider the unitary operator (\ref{rotR}) where $u$ is $\pi $ and $U$ is $-i\Sigma_3 $. Now multiplying by $U^{\dagger}$ on the left and $U$ on the right just changes $\Sigma_1$ and $\Sigma_2$ to $-\Sigma_1$ and $-\Sigma_2$, and does not change $\Sigma_3$. This describes a symmetry of the open dynamics of $S$ when
\begin{eqnarray}
  \label{US1a}
\text{Tr}_R\rho _R[e^{itH}\Sigma_1 e^{-itH}] & = & \Sigma_1 a_1(t) - \Sigma_2 b_1(t), \nonumber \\
\text{Tr}_R\rho _R[e^{itH}\Sigma_2 e^{-itH}] & = & \Sigma_2 a_2(t) + \Sigma_1 b_2(t), \nonumber \\
\text{Tr}_R\rho _R[e^{itH}\Sigma_3 e^{-itH}] & = & \Sigma_3 c(t) + d(t) 
\end{eqnarray}
with functions $a_1(t)$ and $b_1(t)$ in the first equation that may be different from the functions $a_2(t)$ and $b_2(t)$ in the second equation. We have seen\cite{Jordan} that this happens for a number of different combinations of $\gamma_1$, $\gamma_2$, $\gamma_3$ and $\langle \Xi_1 \rangle $, $\langle \Xi_2 \rangle $ and $\langle \Xi_3 \rangle $.

For another set of symmetries, we let $u_1$, $u_2$, $u_3$ be real numbers for which $u_1^{\, 2} + u_2^{\, 2} + u_3^{\, 2}$ is $1$. The one-parameter group of unitary operators
\begin{equation}
\label{rot123} 
U(u) = e^{-iu(1/2)(u_1\Sigma_1 + u_2\Sigma_2 +u_3\Sigma_3)}
\end{equation}
changes the $\Sigma $ matrices by rotation around the axis along the vector $(u_1, u_2, u_3)$ just as the one-parameter group of unitary operators (\ref{rotR}) does for rotation around the $z$ axis. The unitary operators (\ref{rot123}) do not change $u_1\Sigma_1 + u_2\Sigma_2 +u_3\Sigma_3$ and Eqs.(\ref{rot12}) hold when $\Sigma_1$, $\Sigma_2$, $\Sigma_3$ are replaced by 
\begin{eqnarray}
X & = & -u_2\Sigma_1 + u_1\Sigma_2 , \nonumber \\
Y & = & -u_1u_3\Sigma_1 - u_2u_3\Sigma_2 + u_1^{\, 2}\, \Sigma_3 + u_2^{\, 2}\, \Sigma_3, \nonumber \\ G & = & u_1\Sigma_1 + u_2\Sigma_2 +u_3\Sigma_3 ,
\end{eqnarray}
because the commutation relations of $X$ with $G$ and $Y$ with $G$ are the same as the commutation relations of $\Sigma_1$ with $\Sigma_3$ and $\Sigma_2$ with $\Sigma_3$. The vectors $(-u_2, u_1, 0)$ and $(-u_1u_3, -u_2u_3, u_1^{\, 2} + u_2^{\, 2})$ are perpendicular to $(u_1, u_2, u_3)$ and are perpendicular to each other. They are not unit vectors but they have the same length, which is not a factor in Eqs.(\ref{rot12}). The one-parameter group of unitary operators (\ref{rot123}) describes symmetries for the open dynamics of $S$ if Eqs.(\ref{USaa}) hold when $\Sigma_1$, $\Sigma_2$, $\Sigma_3$ are replaced by $X$, $Y$ and $G$. We have calculated, from Eqs.(\ref{USsu})-(\ref{US3}), that this happens when $\gamma_1$, $\gamma_2$, $\gamma_3$ are all equal and the vector $(\langle \Xi_1 \rangle , \langle \Xi_2 \rangle , \langle \Xi_3 \rangle )$ is in the same direction as $(u_1, u_2, u_3)$.

One of the unitary operators (\ref{rot123}) is
\begin{equation}
\label{rotxy} 
U = -i(1/\sqrt{2})(\Sigma_1 + \Sigma_2)
\end{equation}
for rotation by $\pi $ around the axis half way between the $x$ and $y$ axes. It changes $\Sigma_1$, $\Sigma_2$ and $\Sigma_3$ to $\Sigma_2$, $\Sigma_1$ and $-\Sigma_3$. This describes a symmetry of the open dynamics of $S$ if
\begin{eqnarray}
  \label{US1ad}
\text{Tr}_R\rho _R[e^{itH}\Sigma_1 e^{-itH}] & = & \Sigma_1 a_1(t) + \Sigma_2 b_1(t) + \Sigma_3 c_1(t) + d_1(t), \nonumber \\
\text{Tr}_R\rho _R[e^{itH}\Sigma_2 e^{-itH}] & = & \Sigma_1 b_1(t) + \Sigma_2 a_1(t) - \Sigma_3 c_1(t) + d_1(t), \nonumber \\
\text{Tr}_R\rho _R[e^{itH}\Sigma_3 e^{-itH}] & = & \Sigma_1 a_3(t) - \Sigma_2 a_3(t) + \Sigma_3 c_3(t) 
\end{eqnarray}
with functions $a_1(t)$, $b_1(t)$, $c_1(t)$, $d_1(t)$, $a_3(t)$, $c_3(t)$. We can see, from Eqs.(\ref{USsu})-(\ref{US3}), that this happens when $\gamma_1$ and $\gamma_2$ are equal, $\langle \Xi_1 \rangle $ and $\langle \Xi_2 \rangle $ are equal, and $\langle \Xi_3 \rangle $ is zero.

We can, of course, get more examples of symmetries from the ones we have found by making cyclic changes of the indeces $1, 2, 3$. For the Hamiltonians (\ref{Hqd}), we considered all the possibilities for symmetries described by one-parameter groups of unitary operators by looking at Eqs.(\ref{USsu})-(\ref{US3}) to first order in the parameter. Our conclusion is that there are no interesting examples that are significantly different from the ones we have described. We expect that there are interesting examples of discrete symmetries that are significantly different from the ones we have described.

\subsubsection{Hamiltonian $\omega [\alpha \Sigma_2 + \gamma \Sigma_1 \Xi_1] $} \label{Hsecond}

Symmetries occur in a different way when the Hamiltonian is
\begin{equation}
\label{H122}
H = \omega [\alpha \Sigma_2 + \gamma \Sigma_1 \Xi_1]
\end{equation}
with $\alpha$, $\gamma $ and $\omega $ real numbers and $\alpha ^2 + \gamma ^2 = 1$. The one-parameter group of unitary operators
\begin{equation}
\label{UG} 
U(u) = e^{-iu(1/2)G}
\end{equation}
with the generator 
\begin{equation}
\label{G122}
G = \alpha \Sigma_2 + \gamma \Sigma_1
\end{equation}
describes symmetries of the open dynamics of $S$ when there are no correlations between the states of $S$ and $R$ and the state of $R$ is represented by the eigenvector of $\Xi_1 $ for the eigenvalue $1$, because then in the $\text{Tr}_R\rho _R$ in Eq.(\ref{opbasicdep}) the $\Xi_1 $ in $H$ disappears because it commutes with everything that is there and is $1$ in that state, so the $e^{-itH}$ become the same as the $e^{-i(u/2)G}$ and commute with them. The generator $G$ acts as an effective Hamiltonian for $S$; the dynamics in $S$ would be the same if $R$ did not exist and $\omega G$ was the Hamiltonian. The generator $G$ is a dependent constant of the motion for the open dynamics of $S$ as described in Section III.

The results are similar when $\gamma $ is changed to $-\gamma $ in either $H$ or $G$ and the eigenvector of $\Xi_1 $ for the eigenvalue $-1$ represents the state of $R$.

There is another dependent symmetry. When $\langle \Xi_1 \rangle $ is zero, we get
\begin{eqnarray}
  \label{dynforef}
\langle e^{itH}\Sigma_1 e^{-itH}\rangle  & = & \langle \Sigma_1 \rangle \cos^2 \omega t
 + 2\alpha \langle \Sigma_3 \rangle \cos \omega  t\sin \omega  t - (\alpha ^2 - \gamma ^2)\langle  \Sigma_1 \rangle \sin^2 \omega t, 
\nonumber \\
\langle e^{itH}\Sigma_2 e^{-itH}\rangle  & = & \langle \Sigma_2 \rangle \cos^2 \omega t
  + (\alpha ^2 - \gamma ^2)\langle  \Sigma_2 \rangle \sin^2 \omega t,
\nonumber \\
\langle e^{itH}\Sigma_3 e^{-itH}\rangle  & = & \langle \Sigma_3 \rangle \cos^2 \omega t
 - 2\alpha \langle \Sigma_1 \rangle \cos \omega  t\sin \omega  t - \langle  \Sigma_3 \rangle \sin^2 \omega t 
\end{eqnarray}
by using
\begin{equation}
\label{eiH122}
e^{itH} = \cos \omega  t + i[\alpha \Sigma_2 + \gamma \Sigma_1 \Xi_1]\sin \omega  t
\end{equation}
to calculate, for example, that
\begin{eqnarray}
e^{itH}\Sigma_1 e^{-itH} =  \Sigma_1(\cos \omega  t & + & i[-\alpha \Sigma_2 + \gamma \Sigma_1 \Xi_1]\sin \omega  t)\nonumber \\
(\cos \omega  t & - & i[\alpha \Sigma_2 + \gamma \Sigma_1 \Xi_1]\sin \omega  t)
\end{eqnarray}
with the minus sign in front of the $\alpha \Sigma_2$ in the first line coming from taking $\Sigma_1$ through the $\Sigma_2$ from right to left.
From Eqs.(\ref{dynforef}), we can see that $U= \Sigma_2$ describes a dependent symmetry, when $\langle \Xi_1 \rangle $ is zero. It just changes the signs of $\Sigma_1$ and $\Sigma_3$.

\subsubsection{Symmetries outline dynamics structure} \label{symout}

Each different symmetry is characteristic of a different structure for the results of the dynamics. If a symmetry can be assumed, the form it implies for the results of the dynamics can be used, and it may reduce what has to be done to work out the dynamics. For example, if the symmetry described by the unitary operator (\ref{rotxy}) can be assumed, the results of the dynamics can be assumed to have the form of Eqs.(\ref{US1ad}). Of the twelve functions of $t$ that could have been expected as coefficients of $\Sigma_1$, $\Sigma_2$, $\Sigma_3$ and the unit matrix in the three equations, one function is zero and five are repeated, so only six have to be calculated.

When $u$ is $\pi $ and $\alpha$ and $\gamma $ are equal, the unitary symmetry operator (\ref{UG}) is the same as (\ref{rotxy}) and the results of the open dynamics have the form of Eqs.(\ref{US1ad}). The same symmetry and the form it implies for the open dynamics of $S$ come from different Hamiltonians and different states of $R$.

\subsection{One qubit and many qubits} \label{qubits}

Let $S$ be a qubit described by Pauli matrices $\Sigma_1$, $\Sigma_2$, $\Sigma_3$ again and let $R$ be a set of qubits described by Pauli matrices $\Xi^{(k)}_{\; \; 1}$, $\Xi^{(k)}_{\; \; 2}$, $\Xi^{(k)}_{\; \; 3}$. Let
\begin{equation}
\label{Hf}
H = \omega \sum_k (\Sigma_+ \Xi^{(k)}_{\; \; -} + \Sigma_- \Xi^{(k)}_{\; \; +})
\end{equation}
with $\Sigma\pm  = \Sigma_1 \pm i\Sigma_2 $. When there are no correlations between the states of $S$ and $R$ and the state of $R$ is the maximally mixed state, which gives zero for the mean value of every $\Xi^{(k)}_{\; \; 1}$, $\Xi^{(k)}_{\; \; 2}$, $\Xi^{(k)}_{\; \; 3}$, the open dynamics of $S$ has the form of Eqs.(\ref{USaa}) with $b(t)$ and $d(t)$ zero,\cite{Breuer} and the unitary operators (\ref{rotR}) describe a one-parameter group of dependent symmetries. We can write the Hamiltonian (\ref{Hf}) as
\begin{equation}
\label{HSOb}
H = 2\omega \sum_k (\Sigma_1 \Xi^{(k)}_{\; \; 1} + \Sigma_2 \Xi^{(k)}_{\; \; 2}).
\end{equation}
In Section II.A we looked at the case where $R$ is just one qubit and observed that we can get the same result without assuming that the mean value is zero for every $\Xi_1$, $\Xi_2$, $\Xi_3$ if we assume that changes of states for $R$ are symmetries.

\subsection{One oscillator and one oscillator} \label{oscillator}

Let $S$ be an oscillator described by raising and lowering operators $A$ and $A^{\dagger}$ and $R$ an oscillator described by raising and lowering operators $B$ and $B^{\dagger}$ so
\begin{equation}
\label{ABcom}
[A,A^{\dagger}] = 1, \, \, \, \, [B,B^{\dagger}] = 1,
\end{equation}
and $A$ and $A^{\dagger}$ commute with $B$ and $B^{\dagger}$, as in Section II.D of the preceding paper.\cite{Jordan}

\subsubsection{One-parameter group of symmetries} \label{group}

We will consider two different Hamiltonians. The first is
\begin{eqnarray}
\label{Hoo}
H & = & \frac{\omega }{2}(A + B)^{\dagger}(A + B)
 \nonumber \\ & + & \frac{\eta }{2}(A - B)^{\dagger}(A - B).
\end{eqnarray}
It gives
\begin{eqnarray}
\label{A+Bt}
e^{itH}(A + B)e^{-itH} & = & (A + B)e^{-i\omega t} \nonumber \\
e^{itH}(A - B)e^{-itH} & = & (A - B)e^{-i\eta t}
\end{eqnarray}
\begin{eqnarray}
\label{AorAdagt}
e^{itH}Ae^{-itH} & = & A\frac{1}{2}(e^{-i\omega t} + e^{-i\eta t}) + B\frac{1}{2}(e^{-i\omega t} - e^{-i\eta t})\nonumber \\
e^{itH}A^{\dagger}e^{-itH} & = & A^{\dagger}\frac{1}{2}(e^{i\omega t} + e^{i\eta t}) + B^{\dagger}\frac{1}{2}(e^{i\omega t} - e^{i\eta t}).
\end{eqnarray}

There are dependent symmetries described by the one-parameter group of operators
\begin{equation}
\label{U(u)}
U(u) = e^{-iuA^{\dagger}A}
\end{equation}
for real $u$. They change $A$ to 
\begin{equation}
\label{UonA}
U(u)^{\dagger}AU(u) = Ae^{-iu}.
\end{equation}
They change $A^{\dagger}$ to $A^{\dagger}$ multiplied by $e^{iu}$. They do not change $B$ and $B^{\dagger}$.

These are symmetries of the open dynamics of $S$ when there are no correlations between the states of $S$ and $R$ and the state of $R$ is represented by an eigenvector of $B^{\dagger}B$. Every operator $Q$ for $S$ is a function of $A$ and $A^{\dagger}$. The $e^{itH}$ and $e^{-itH}$ change it to the same function of the operators described by Eqs.(\ref{AorAdagt}). On the right side of Eq.(\ref{opbasicdep}), where $U^{\dagger}$ and $U$ act after $e^{itH}$ and $e^{-itH}$, they change only the $A$ and $A^{\dagger}$. They do not change the $B$ and $B^{\dagger}$ that are brought in by Eqs.(\ref{AorAdagt}). On the left side of Eq.(\ref{opbasicdep}), the $B$ and $B^{\dagger}$ are brought in multiplied by the $e^{-iu}$ and $e^{iu}$ that the $U^{\dagger}$ and $U$ put on the $A$ and $A^{\dagger}$ before the $e^{itH}$ and $e^{-itH}$ act. If these phases on the $B$ and $B^{\dagger}$ cancel out, the two sides of Eq.(\ref{opbasicdep}) are the same. They do cancel out when the state of $R$ is represented by an eigenvector of $B^{\dagger}B$; then the mean value of a product of powers of $B$ and powers of $B^{\dagger}$ is zero unless the number of $B$ is the same as the number of $B^{\dagger}$.

\subsubsection{Discrete symmetry} \label{discrete}

Now we consider the Hamiltonian
\begin{eqnarray}
\label{Hoo2}
H & = & i\frac{\omega }{2}[(A + B)^{\dagger}(A + B)^{\dagger} - (A + B)(A + B)]
 \nonumber \\ & + & i\frac{\eta }{2}[(A - B)^{\dagger}(A - B)^{\dagger} - (A - B)(A - B)].
\end{eqnarray}
It gives
\begin{eqnarray}
\label{A+Bt2}
e^{itH}(A + B)e^{-itH} & = & (A + B)\cosh\omega t + (A^{\dagger} + B^{\dagger})\sinh\omega t \nonumber \\
e^{itH}(A - B)e^{-itH} & = & (A - B)\cosh\eta t + (A^{\dagger} - B^{\dagger})\sinh\eta t
\end{eqnarray}
\begin{eqnarray}
\label{AorAdagt2}
e^{itH}Ae^{-itH} & = & A\frac{1}{2}(\cosh\omega t + \cosh\eta t) + A^{\dagger}\frac{1}{2}(\sinh\omega t + \sinh\eta t) \nonumber \\
& + & B\frac{1}{2}(\cosh\omega t - \cosh\eta t) + B^{\dagger}\frac{1}{2}(\sinh\omega t - \sinh\eta t) \nonumber \\
e^{itH}A^{\dagger}e^{-itH} & = & A^{\dagger}\frac{1}{2}(\cosh\omega t + \cosh\eta t) + A\frac{1}{2}(\sinh\omega t + \sinh\eta t) \nonumber \\
& + & B^{\dagger}\frac{1}{2}(\cosh\omega t - \cosh\eta t) + B\frac{1}{2}(\sinh\omega t - \sinh\eta t).
\end{eqnarray}

In place of the one-parameter group of symmetries described by the operators $e^{-iuA^{\dagger}A}$, we now have a discrete symmetry described by
\begin{equation}
\label{sgnch}
U = e^{-i\pi A^{\dagger}A}
\end{equation}
which gives
\begin{equation}
\label{UonA2}
U(u)^{\dagger}AU(u) = -A, \;\;\;\;\; U(u)^{\dagger}A^{\dagger}U(u) = -A^{\dagger}.
\end{equation}
It describes a symmetry of the open dynamics of $S$ for the Hamiltonian (\ref{Hoo2}), when there are no correlations between the states of $S$ and $R$ and the state of $R$ is represented by an eigenvector of $B^{\dagger}B$, just as the $e^{-iuA^{\dagger}A}$ did for the Hamiltonian (\ref{Hoo}).

\subsubsection{Symmetries distinguish forms of dynamics results} \label{saywhat}

These symmetries provide substantial statements about what the results of the dynamics can be. The one-parameter group of symmetries that change $A$ to $Ae^{-iu}$ is compatible with results of the form
\begin{equation}
  \label{result}
\text{Tr}_R\rho _R[e^{itH}Ae^{-itH}] = Af(t) 
\end{equation}
that we get from the Hamiltonian (\ref{Hoo}) when there are no correlations and the state of $R$ is represented by an eigenvector of $B^{\dagger}B$, but only the discrete symmetry that changes $A$ to $-A$ and $A^{\dagger}$ to $-A^{\dagger}$ is compatible with results of the form
\begin{equation}
  \label{result}
\text{Tr}_R\rho _R[e^{itH}Ae^{-itH}] = Af(t) + A^{\dagger}g(t) 
\end{equation}
that we get from the Hamiltonian (\ref{Hoo2}) when there are no correlations and the state of $R$ is represented by an eigenvector of $B^{\dagger}B$, and none of these symmetries are compatible with results of the form
\begin{equation}
  \label{result}
\text{Tr}_R\rho _R[e^{itH}Ae^{-itH}] = Af(t) + A^{\dagger}g(t) + b(t)
\end{equation}
that we would get if the mean value $\langle B\rangle $ were not zero for the state of $R$.

\subsubsection{Classical mechanics} \label{classical}

The examples of this Section II.C can be translated into classical mechanics quite simply. In terms of
\begin{eqnarray}
\label{Qs&Ps}
Q_A & = & \frac{1}{\sqrt{2}}(A + A^{\dagger}), \,\,\,\, P_A = \frac{1}{i\sqrt{2}}(A - A^{\dagger}),\nonumber \\
Q_B & = & \frac{1}{\sqrt{2}}(B + B^{\dagger}), \,\,\,\, P_B = \frac{1}{i\sqrt{2}}(B - B^{\dagger})
\end{eqnarray}
the dynamics generated by the Hamiltonian (\ref{Hoo}) and described by Eqs.(\ref{A+Bt}) and (\ref{AorAdagt}) is that $Q_A$, $P_A$, $Q_B$, $P_B$ are changed to
\begin{eqnarray}
\label{tQs&Ps}
Q_A(t) & = & Q_A\frac{1}{2}(\cos \omega  t + \cos \eta  t) + P_A\frac{1}{2}(\sin \omega  t + \sin \eta  t) \nonumber \\
& + & Q_B\frac{1}{2}(\cos \omega  t - \cos \eta  t) + P_B\frac{1}{2}(\sin \omega  t - \sin \eta  t),\nonumber \\
P_A(t) & = & P_A\frac{1}{2}(\cos \omega  t + \cos \eta  t) - Q_A\frac{1}{2}(\sin \omega  t + \sin \eta  t) \nonumber \\
& + & P_B\frac{1}{2}(\cos \omega  t - \cos \eta  t) - Q_B\frac{1}{2}(\sin \omega  t - \sin \eta  t),\nonumber \\
Q_B(t) & = & Q_B\frac{1}{2}(\cos \omega  t + \cos \eta  t) + P_B\frac{1}{2}(\sin \omega  t + \sin \eta  t) \nonumber \\
& + & Q_A\frac{1}{2}(\cos \omega  t - \cos \eta  t) + P_A\frac{1}{2}(\sin \omega  t - \sin \eta  t),\nonumber \\
P_B(t) & = & P_B\frac{1}{2}(\cos \omega  t + \cos \eta  t) - Q_B\frac{1}{2}(\sin \omega  t + \sin \eta  t) \nonumber \\
& + & P_A\frac{1}{2}(\cos \omega  t - \cos \eta  t) - Q_A\frac{1}{2}(\sin \omega  t - \sin \eta  t)
\end{eqnarray}
at time $t$. The dynamics generated by the Hamiltonian (\ref{Hoo2}) and described by Eqs.(\ref{A+Bt2}) and (\ref{AorAdagt2}) is that $Q_A$, $P_A$, $Q_B$, $P_B$ are changed to
\begin{eqnarray}
\label{tQ&Ps}
Q_A(t) & = & Q_A\frac{1}{2}(\cosh \omega  t + \cosh \eta  t + \sinh \omega  t + \sinh \eta  t) \nonumber \\
& + & Q_B\frac{1}{2}(\cosh \omega  t - \cosh \eta  t + \sinh \omega  t - \sinh \eta  t),\nonumber \\
P_A(t) & = & P_A\frac{1}{2}(\cosh \omega  t + \cosh \eta  t - \sinh \omega  t - \sinh \eta  t) \nonumber \\
& + & P_B\frac{1}{2}(\cosh \omega  t - \cosh \eta  t - \sinh \omega  t + \sinh \eta  t),\nonumber \\
Q_B(t) & = & Q_B\frac{1}{2}(\cosh \omega  t + \cosh \eta  t + \sinh \omega  t + \sinh \eta  t) \nonumber \\
& + & Q_A\frac{1}{2}(\cosh \omega  t - \cosh \eta  t + \sinh \omega  t - \sinh \eta  t),\nonumber \\
P_B(t) & = & P_B\frac{1}{2}(\cosh \omega  t + \cosh \eta  t - \sinh \omega  t - \sinh \eta  t) \nonumber \\
& + & P_A\frac{1}{2}(\cosh \omega  t - \cosh \eta  t - \sinh \omega  t + \sinh \eta  t)
\end{eqnarray}
at time $t$. The one-parameter group of transformations that multiply $A$ by $e^{-iu}$ and $A^{\dagger}$ by $e^{iu}$, as described by Eqs.(\ref{U(u)}) and (\ref{UonA}), is that $Q_A$, $P_A$ are changed to
\begin{equation}
  \label{QPu}
Q_A(u) = Q_A\cos u + P_A\sin u, \,\,\,\, P_A(u) = P_A\cos u - Q_A \sin u.
\end{equation}
In the special case when $u$ is $\pi $ and $A$ and $A^{\dagger}$ are changed to $-A$ and $-A^{\dagger}$, as described in Eqs.(\ref{sgnch}) and (\ref{UonA2}), $Q_A$ and $P_A$ are changed to $-Q_A$ and $-P_A$. All of these are canonical transformations of canonical coordinates and momenta $Q_A$, $P_A$, $Q_B$, $P_B$.

The physical quantities for the subsystem $S$ are functions of $Q_A$ and $P_A$. The requirement for a symmetry of the open classical dynamics of $S$ like the requirement of Eq.(\ref{mvbasic}) for a symmetry of the open quantum dynamics, is that the mean value of a function of $Q_A$ and $P_A$ that is changed first by the canonical transformation for the symmetry and then by a canonical transformation for the dynamics is the same as when it is changed first by the canonical transformation for the dynamics and then by the canonical transformation for the symmetry.

Mean values are calculated with a probability density function that describes the state of the classical system. For the examples considered here, it is assumed that there are no correlations between the states of $S$ and $R$, so the probability density function for the entire system of $S$ and $R$ combined is a product $\rho_S (Q_A, P_A)\rho_R (Q_B, P_B)$ of a probability density function of $Q_A$ and $P_A$ for $S$ and a probability density function of $Q_B$ and $P_B$ for $R$. A symmetry is assumed to be for all the states of $S$, so the requirement for a symmetry must hold for all probability density functions $\rho_S (Q_A, P_A)$. That means it must hold for the individual values that are averaged with the probabilities $\rho_S (Q_A, P_A)$ to get the mean values. The symmetry requirement can be stated with the mean value for the state of $S$ removed. As in Eq.(\ref{opbasicdep}), only the mean value for the state of $R$ remains. For a dependent symmetry, the requirement will hold for particular states of $R$ described by particular probability density functions $\rho_R (Q_B, P_B)$.

The rotations of $Q_A$ and $P_A$ described by Eqs.(\ref{QPu}) are symmetries of the dynamics described by Eqs.(\ref{tQs&Ps}) for particular states of $B$. When $Q_A$ and $P_A$ are changed first by the dynamics and then by a rotation, the overall change of $Q_A$ and $P_A$ is to the $Q_A (t)$ and $P_A (t)$ of Eqs.(\ref{tQs&Ps}) in which the $Q_A$ and $P_A$ are rotated as in Eqs.(\ref{QPu}). When $Q_A$ and $P_A$ are changed first by a rotation and then by the dynamics, the overall change of $Q_A$ and $P_A$ is to the $Q_A (t)$ and $P_A (t)$ of Eqs.(\ref{tQs&Ps}) in which the $Q_A$ and $P_A$ are rotated as in Eqs.(\ref{QPu}) and the $Q_B$ and $P_B$ are rotated the same as the $Q_A$ and $P_A$. The requirement for a symmetry holds if mean values for the state of $R$ are not changed by rotations of $Q_B$ and $P_B$. This happens when $\rho_R (Q_B, P_B)$ is a function only of $Q_B^{\, \, \, 2} + P_B^{\, \, \, 2}$.

The change of $Q_A$ and $P_A$ to $-Q_A$ and $-P_A$ is a symmetry of the dynamics described by Eqs.(\ref{tQ&Ps}) for particular states of $B$. When the dynamics is first and the sign change second, the overall change of $Q_A$ and $P_A$ is to the $Q_A (t)$ and $P_A (t)$ of Eqs.(\ref{tQ&Ps}) in which the $Q_A$ and $P_A$ are changed to $-Q_A$ and $-P_A$. When the sign change is first and the dynamics second, the overall change of $Q_A$ and $P_A$ is to the $Q_A (t)$ and $P_A (t)$ of Eqs.(\ref{tQ&Ps}) in which the $Q_A$ and $P_A$ are are changed to $-Q_A$ and $-P_A$ and the $Q_B$ and $P_B$ are changed to $-Q_B$ and $-P_B$. The requirement for a symmetry holds if mean values for the state of $R$ are not changed when $Q_B$ and $P_B$ are changed to $-Q_B$ and $-P_B$. This happens when $\rho_R (Q_B, P_B)$ is a function only of $Q_B^{\, \, \, 2} $, $P_B^{\, \, \, 2}$ and $Q_BP_B$.

\subsection{One qubit and one oscillator} \label{qubit and oscillator}

Let $S$ be a qubit described by Pauli matrices $\Sigma_1$, $\Sigma_2$, $\Sigma_3$, as in Section II.A, and $R$ an oscillator described by raising and lowering operators $B$ and $B^{\dagger}$, as in Section II.C, and let
\begin{equation}
\label{HSO}
H = \omega (\Sigma_+B + \Sigma_-B^{\dagger})
\end{equation}
with $\Sigma\pm  = \Sigma_1 \pm i\Sigma_2 $ as before. This gives
\begin{eqnarray}
  \label{US1}
e^{itH}\Sigma_1 e^{-itH} & = & \Sigma_1 e^{i\omega t (\Sigma_- B + \Sigma_+ B^{\dagger})}e^{-i\omega t (\Sigma_+ B + \Sigma_- B^{\dagger})}, \nonumber \\
e^{itH}\Sigma_2 e^{-itH} & = & \Sigma_2 e^{-i\omega t (\Sigma_- B + \Sigma_+ B^{\dagger})}e^{-i\omega t (\Sigma_+ B + \Sigma_- B^{\dagger})}, \nonumber \\
e^{itH}\Sigma_3 e^{-itH} & = & \Sigma_3 e^{-i2\omega t (\Sigma_+ B + \Sigma_- B^{\dagger})}. 
\end{eqnarray}

The one parameter group of rotation operators
\begin{equation}
\label{rotR2} 
U(u) = e^{-iu(1/2)\Sigma_3}
\end{equation}
for real $u$ represent symmetries of the open dynamics of $S$ when, as in Section II.C, there are no correlations between the states of $S$ and $R$ and the state of $R$ is represented by an eigenvector of $B^{\dagger}B$. Then, from Eq.(\ref{US1}),
\begin{eqnarray}
  \label{US1a2}
\text{Tr}_R\rho _R[e^{itH}\Sigma_1 e^{-itH}] & = & \Sigma_1 [a(t) + b(t)\Sigma_3 ] = \Sigma_1 a(t) - \Sigma_2 b(t), \nonumber \\
\text{Tr}_R\rho _R[e^{itH}\Sigma_2 e^{-itH}] & = & \Sigma_2 [a(t) + b(t)\Sigma_3 ] = \Sigma_2 a(t) + \Sigma_1 b(t), \nonumber \\
\text{Tr}_R\rho _R[e^{itH}\Sigma_3 e^{-itH}] & = & \Sigma_3 [c(t) + d(t)\Sigma_3 ] = \Sigma_3 c(t) + d(t) 
\end{eqnarray}
with real functions $a(t)$, $b(t)$, $c(t)$, $d(t)$, because in each of the products of powers of $B$ and powers of $B^{\dagger}$ for which the $\text{Tr}_R\rho _R $ does not give zero, the number of $B$ must be the same as the number of $B^{\dagger}$, as in Section II.C, so each of the accompanying products of powers of $\Sigma_+$ and $\Sigma_-$ must contain an even number of factors and be a product of pairs
\begin{eqnarray}
  \label{sigprod}
\Sigma_+ \Sigma_+  & = & 0, \nonumber \\
\Sigma_- \Sigma_-  & = & 0, \nonumber \\
\Sigma_+ \Sigma_-  & = & 2 + 2\Sigma_3 ,\nonumber \\
\Sigma_- \Sigma_+  & = & 2 - 2\Sigma_3 . 
\end{eqnarray}
The functions $a(t)$ and $b(t)$ are the same in the second as in the first of Eqs.(\ref{US1a}) because the minus sign that is in the second but not the first of Eqs.(\ref{US1}) cancels out: it cancels out in a product where the number of $B$ and $B^{\dagger}$ factors that come from each of the $e^{-i\omega t( \; \; )}$ factors in the second of Eqs.(\ref{US1}) is even; if the number of $B$ and $B^{\dagger}$ factors that come from each of the $e^{-i\omega t( \; \; )}$ factors in the second of Eqs.(\ref{US1}) is odd, then either there are at least two more $\Sigma_+ $ factors than $\Sigma_-$ factors and a pair $\Sigma_+ \Sigma_+ $ that gives zero or there are at least two more $\Sigma_- $ factors than $\Sigma_+ $ factors and a pair $\Sigma_- \Sigma_- $ that gives zero.

The symmetry operators do not commute with $H$, but the symmetry generator $\Sigma_3$ represents  a quantity that is is a constant of the motion for the open dynamics of $S$ when the state of $R$ is represented by the eigenvector of $B^{\dagger}B$ for the eigenvalue zero. The significance of this example is put in question by the fact that, for that state of $R$, every function of $\Sigma_1$, $\Sigma_2$, $\Sigma_3$ represents a constant of the motion as well as $\Sigma_3$.

The symmetries for the open dynamics of $S$ described by the unitary operators (\ref{rotR2}) and the end form of Eqs.(\ref{US1a2}) are the same as for Eqs.(\ref{rotR}) and (\ref{US1a}) of Section II.A.

\subsection{One qubit and many oscillators} \label{qubit and oscillators}

Let $S$ be a qubit described by Pauli matrices $\Sigma_1$, $\Sigma_2$, $\Sigma_3$, as before, and let $R$ a number of  oscillators, described by raising and lowering operators $B_k$ and $B_k^{\, \dagger}$, so the operators $B_k$ and $B_k^{\, \dagger}$ commute with each other for different $k$ and
\begin{equation}
\label{Bcom}
[B_k, B_k^{\, \dagger}] = 1
\end{equation}
for each $k$. Let
\begin{equation}
\label{HSO}
H = \omega \sum_k (\Sigma_+ B_k + \Sigma_- B_k^{\, \dagger})
\end{equation}
with $\Sigma\pm  = \Sigma_1 \pm i\Sigma_2 $ as before. When there are no correlations between the states of $S$ and $R$ and the state of $R$ is represented by an eigenvector of $B_k^{\, \dagger}B_k $ with eigenvalue zero for every $k$, the open dynamics of $S$ has the form of Eqs.(\ref{US1a}) again and the unitary operators (\ref{rotR}) again describe a one-parameter group of dependent symmetries.\cite{Morozov}

\section{Constants of the motion}\label{three}

Are symmetries related to constants of the motion in open quantum dynamics? How are constants of the motion defined in open quantum dynamics? If we think about constants of the motion the same way we think about independent symmetries, and consider a statement that an operator $Q$ for $S$ represents a quantity that is a constant of the motion for the open dynamics of $S$, we can say\cite{Jordan} that it must mean that $Q$ commutes with the Hamiltonian $H$ for the dynamics of the entire system of $S$ and $R$ combined. In particular, if $Q$ is a unitary symmetry operator, or an Hermitian operator that is a generator of a one-parameter group of symmetry operators, for independent symmetries, we would say that $Q$ can represent a constant of the motion for the open dynamics of $S$ only if $Q$ commutes with $H$, which means that it describes a symmetry for the dynamics of the entire system of $S$ and $R$ combined.

Here we will think about constants of the motion the same way we think about dependent symmetries. We will say that an operator $Q$ for $S$ represents a quantity that is a \textit{dependent constant of the motion} for the open dynamics of $S$ if it is constant for all possible initial states of $S$ but only for particular states of $R$ or correlations, or absence of correlations, between the states of $S$ and $R$. If we assume there are no correlations, we will say that Q represents a dependent constant of the motion if
\begin{equation}
\label{constantd}
\text{Tr}_R \rho _R \left[e^{itH}f(Q)e^{-itH}\right] = f(Q)
\end{equation}
for all $t$, for any functions $f(Q)$, but only for particular states of $R$. If $Q$ is a unitary symmetry operator, or an Hermitian operator that is a generator of a one-parameter group of symmetry operators, for dependent symmetries for particular states of $R$, we can ask whether $Q$ represents a dependent constant of the motion for those same states of $R$. 

Using powers of $Q$ for the $f(Q)$, or the projection operators from the spectral decomposition of $Q$, would make a statement that the probabilities for values of the quantity represented by $Q$ are constant.  If $S$ is a qubit, a physical quantity for $S$ has no more than two possible values, so the probabilities for its values are constant if its mean value is constant. All that is needed from Eqs.(\ref{constantd}) is that 
\begin{equation}
\label{constantmvd}
\text{Tr}_R \rho _R \left[e^{itH}Qe^{-itH}\right] = Q.
\end{equation}
In the examples we will consider now, $S$ will be a qubit described by Pauli matrices $\Sigma_1$, $\Sigma_2$, $\Sigma_3$ and $R$ a qubit described by Pauli matrices $\Xi_1$, $\Xi_2$, $\Xi_3$ as in Section II.A. We will assume that there are no correlations between the states of $S$ and $R$.

\subsection{Sometimes nothing is constant} \label{nothing}

For the Hamiltonian (\ref{Hqd}) with $\gamma_1$, $\gamma_2$, $\gamma_3$ all equal, we can see that the $\cos^2 \gamma t $ in the first term on the right in each of Eqs.(\ref{USsu})-(\ref{US3}) makes it impossible for Eq.(\ref{constantmvd}) to hold for any state of $R$ for any $Q$ other than a multiple of the identity operator. No quantity that can have two different values can be a dependent constant of the motion for this Hamiltonian. In particular, we see two symmetry generators that can not be dependent constants of the motion. The one-parameter groups of unitary operators (\ref{rotR}) and (\ref{rot123}) both describe symmetries of the open dynamics of $S$ for this Hamiltonian for particular states of $R$. Their generators $\Sigma_3$ and $u_1\Sigma_1 + u_2\Sigma_2 +u_3\Sigma_3$ do not represent dependent constants of the motion. 

Another example is described at the end of Section II.D.

\subsection{Sometimes everything is constant} \label{everything}

Suppose the Hamiltonian is
\begin{equation}
\label{Heverything}
H = \omega (\Sigma_2 - \Sigma_2 \Xi_2 )
\end{equation}
with $\omega $ a real number, there are no correlations between the states of $S$ and $R$, and the state of $R$ is represented by the eigenvector of $\Xi_2 $ for the eigenvalue $1$. The result of this combination is similar to that of Section II.A.2. In the $\text{Tr}_R\rho _R$ in Eq.(\ref{constantmvd}) the $\Xi_2 $ in $H$ disappears, because it commutes with everything that is there and is $1$ in that state of $R$, so the Hamiltonian (\ref{Heverything}) becomes zero and Eq.(\ref{constantmvd}) holds for all $Q$ for $S$. Every quantity for $S$ is a dependent constant of the motion.

\subsection{A symmetry generator can be constant} \label{symgencon}

The symmetry generator $G$ of Eq.(\ref{G122}) in Section II.A.2 represents a dependent constant of the motion for the open dynamics of $S$ for the Hamiltonian (\ref{H122}) when there are no correlations between the states of $S$ and $R$ and the state of $R$ is represented by the eigenvector of $\Xi_1 $ for the eigenvalue $1$, because then in the $\text{Tr}_R\rho _R$ in Eq.(\ref{constantmvd}) the $\Xi_1 $ in $H$ disappears because, again, it commutes with everything that is there and is $1$ in that state of $R$, so $H$ becomes the same as $\omega G$ and commutes with $G$.


\begin{thebibliography}{12}
\bibitem{note}The equations also reformulate the rule slightly for the symmetries of the dynamics of an entire closed system: when the spectrum of the Hamiltonian does not have a lower bound, they allow a phase factor which does not change the physical meaning.
\end{thebibliography}
\end{document}